\documentclass[twocolumn,showpacs,preprintnumbers,amsmath,amssymb]{revtex4-1}
\usepackage{graphicx}
\usepackage{dcolumn}
\usepackage{siunitx}
 \usepackage[cmyk]{xcolor}
 \usepackage{latexsym}
 \usepackage{mathrsfs}
\usepackage[abx]{MnSymbol}
\usepackage{gensymb}
\usepackage{tikz}
\usepackage{eucal}
\usepackage[version=3]{mhchem}
\usepackage{amssymb}
\usepackage{float}
\usepackage{siunitx}

\begin{document}

\title{Controlling light emission by  engineering atomic geometries in silicon photonics}

\author{Arindam Nandi$^2$, Xiaodong Jiang$^2$,  Dongmin Pak$^2$, Daniel Perry$^1$, Kyunghun Han $^{2}$, Edward S Bielejec$^1$, Yi Xuan$^2$, Mahdi Hosseini$^{2,3,*}$}

\affiliation{$^1$ Sandia National Laboratory Albuquerque NM 87185, USA\\
$^2$ Birck Nanotechnology Center, Department of Electrical and Computer Engineering, Purdue University West Lafayette IN 47907, USA\\
$^3$ Purdue Quantum Science and Engineering Institute, Purdue University, West Lafayette, Indiana 47907, USA\\
$^*$ mh@purdue.edu}

%\date{\today}

\begin{abstract} %This format begins with an introductory paragraph (not abstract) of 150 words maximum, summarizing the background, rationale, main results and implications. This paragraph should be referenced, as in Nature style, and should be considered part of the main text, so that any subsequent introductory material avoids too much repetition of the introductory paragraph.

%Controlling intermodal coupling between multiple excitations within a photonic resonator may enable design of novel quantum photonic metamaterials exhibiting anomalous effects. 

By engineering atomic geometries composed of nearly 1000 atomic segments embedded in micro-resonators, we observe Bragg resonances induced by the atomic lattice at the telecommunication wavelength. The geometrical arrangement of erbium atoms into a lattice inside a silicon nitride microring resonator reduces the scattering loss at a wavelength commensurate with the lattice. We confirm dependency of light emission to the atomic positions and lattice spacing and also observe Fano interference between resonant modes in the system.

\end{abstract}

\maketitle

Novel phenomena emerge when resonant modes of a hybrid system undergo coherent interactions~\cite{limonov2017fano}. The interactions of this kind may result in engineering unconventional materials and platforms for broad applications. In photonics, the coherent and cooperative mode coupling has resulted in observation of peculiar effects including Fano interference~\cite{limonov2017fano}, cooperative light scattering,~\cite{goban2015superradiance,sorensen2016coherent, Corzo:2016aa}, cavity quantum electrodynamic (cQED) interactions~\cite{Roy:2017aa,Haroche:2013aa}, Borrmann effect~\cite{borrmann1941extinktionsdiagramme, novikov2017borrmann}, and topological  optical effects~\cite{Lu:2014aa, Barik:2018aa, mittal2018topological}.    Also, controlling the position of laser-cooled atoms near waveguides  and fibers has led to the observation of peculiar effects such as coherent backscattering\cite{sorensen2016coherent}and superradiance\cite{goban2015superradiance}. In a different platform, photon-mediated coupling between a small ensemble of microwave oscillators in a superconducting circuit has been observed\cite{Mirhosseini:nature2019}.  
In solid photonics,  engineering light-atom interactions to achieve long-range coherent interference in the medium has not been achieved due to the lack of control on atomic positioning. Towards realization of scalable and long-range interference between fields in an atomic ensemble, we study the effect of atomic geometry on photon emission in solid-state photonics. The atoms, in this case, are erbium ions which are rare earth ions with a telecom-band transition embedded in silicon nitride materials. The approach enables study of a novel regime of light-matter interactions in solids by designing the atomic geometry and mode coupling in the system.  
The relatively low sensitivity of the rare earth (RE) ions to the solid's environment makes these ions an attractive substance for realization of linear and nonlinear light-matter interactions for quantum applications\cite{Hedges:2010p11910,Erhan:nphot2015,Dibos:2018aa}. 

Here, by activation of silicon nitride structures using precision  implantation of isotopically pure Er ions, we study coupling between optical and atomic \color{black} Bragg \color{black} resonant modes.  We observe emission \color{black} enhancement \color{black} when the embedded atomic lattice is commensurate with the wavelength of the emitted light. This is because, the emitted light from the atomic lattice gives rise to coherent spatial interference creating a Bragg atomic resonance. On the resonant condition where the emission wavelength is commensurate with the atomic lattice, the \color{black} scattering loss decreases. \color{black} Moreover, we record asymmetric  lineshapes of photon emission governed by Fano interference between the resonant modes of the system.  In this way, we take advantage of the inhomogeneous broadening of ions in silicon nitride to enhance \color{black} collection of radiation \color{black} at a desired wavelength \color{black} within the emission band of the Er ions\color{black}.

%The ultimate example of such effect was given by Borrmann in 1941~\cite{borrmann1941extinktionsdiagramme} describing anomalous transmission of x-ray beam though perfect crystals. 

\begin{figure}[!t]
\vspace{-1mm}
	\centerline{\includegraphics[width=0.9\columnwidth, angle=0]{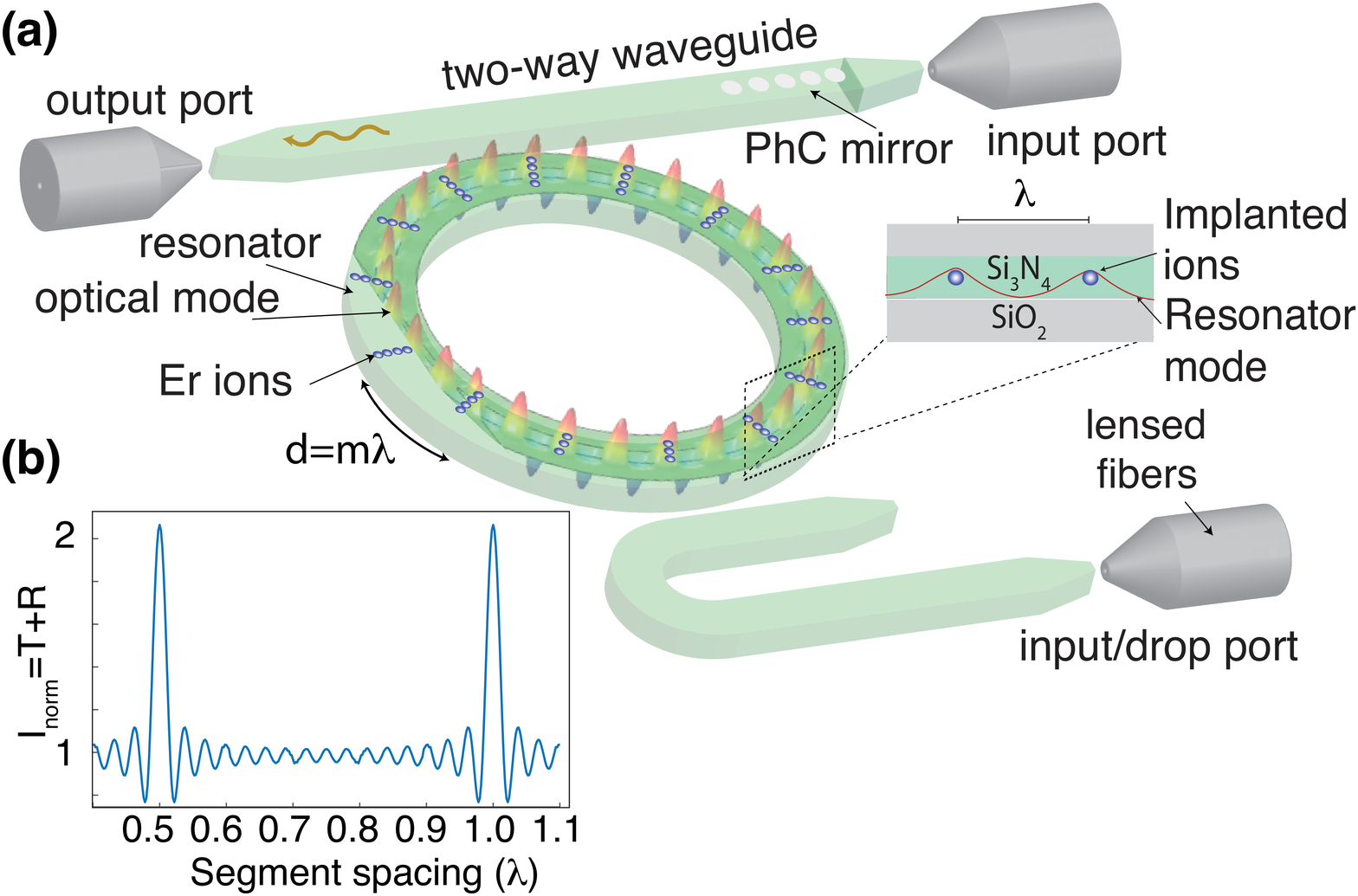}}
	\vspace{0mm}
	\caption{{\bf Schematic of the designed active SiN ring resonator.} (a)Isotopically pure $^{168}$Er ions are implanted as an array inside a microring cavity. The segments are separated by multiples of the wavelength of Er emission. At each segment with a width much smaller than the wavelength, around $10^{4}$ ions were implanted along the diameter of the ring. A photonic crystal mirror is used to increase the collection efficiency to the lensed fiber on the left (output port). The excitation pump enters the ring either from the drop-port or the mirror-side waveguide using another lensed fiber. (b) Result of numerical simulation  \color{black} for total field intensity ($T+R=|\mathcal{E}_b(z=0)|^2 + |\mathcal{E}_f(z=L)|^2$) as a function of atom spacing  from a lattice of atoms, which is normalized to that of a random atomic distribution. \color{black} In this simulation, 15 atomic segments were considered. \color{black}}
	\label{fig: fig1}
\vspace{-5mm}\end{figure}

 We use isotopically pure $^{168}$Er ions directly implanted in a SiN microring resonator into a periodic array (atomic lattice) (see Fig.\ref{fig: fig1}(a) ).  Considering an array of ion segments (with the width of each segment being much smaller than the wavelength) implanted in a ring resonator, the emission from the ion array can be approximately described by a 1D theory of atoms coupled to a waveguide. We  study the interaction dynamics by numerically solving the semiclassical Maxwell-Bloch equations of motion. The equations of motion capturing the interaction of light with an ensemble of atoms in free space are
\begin{eqnarray}
\hspace{-5mm}\frac{d\mathcal{E}_{b/f}}{dz} &=& \pm i\mathcal{N}(z)\sigma_{12} \\
\hspace{-5mm}\frac{d\sigma_{12}}{dt} &=& -(\gamma_h+i \delta\omega) \sigma_{12} \color{black} + \color{black} ig (\mathcal{E}_f^*+\mathcal{E}_be^{2ikz}) (\sigma_{22}-\sigma_{11})
\end{eqnarray}
where $\mathcal{E}_{f/b}$ is the expected value of the forward/backward intra-cavity electric field operator, $\mathcal{N}(z)$ is a function describing the linear atomic density, $\sigma_{12}$ is the amplitude of the atomic polarization in a two-level atom,$\gamma_h$ is the docoherence rate which includes both population decay rate ($\gamma=1/T_1$) and excess dephasing (\color{black}$\gamma_p$\color{black}) and $g$ is the light-atom coupling rate. The inhomogeneous broadening of the atomic transition is described by $\delta\omega$. We numerically solve these equations for various atom spacing. The result for emission from the array is shown in Fig.\ref{fig: fig1}(b) as a function of the atomic spacing. The maximum \color{black}intensity is observed \color{black}  when the atom spacing is a multiple of half of the resonant wavelength. This is consistent with the free-space Bragg theory\cite{DEUTSCH:1995aa}. \color{black} In the case of a continuously driven ring resonator  and in the limit of $\kappa>\gamma$, the cavity field with decay rate $\kappa$ can be adiabatically eliminated with light-atom dynamics approximately described by Eqs.(1)-(2). The mode mixing \cite{Chang:2019aa, Feng:2014ab} and coherent interference in the spatially ordered lattice \cite{DEUTSCH:1995aa} in the ring resonator gives rise to a standing wave structure for the scattered light \cite{Slama:2006aa}. The atomic lattice acts as a Bragg grating \cite{sorensen2016coherent} scattering light in both directions  with probability \color{black} $|\beta_{eff}/(1+\beta_{eff})|^2$ \color{black} , where $\beta_{eff}$ is the \color{black} effective \color{black} ratio of scattering into the resonator mode \color{black} by the ensemble of the atomic segments \cite{furuya2020study}\color{black}, $N_{eff}g^2/\kappa$, to that of free space, $\gamma$, and $N_{eff}$ is the effective atom number contributing to the scattering process. Compared to the free-space case, $\beta_{eff}$ is enhanced in the ring by a factor of $F/\pi$ \cite{Haruka:AAMOP2011}, where $F$ is the \color{black} equivalent \color{black} resonator finesse. With standing wave nodes of the scattered light at the position of the atoms, the scattering loss is expected to reduce\cite{DEUTSCH:1995aa}. We note that  reduction in scattering occurs in all directions due to lower field intensity at the location of the atoms. \color{black} This is different from Purcell enhancement or superradiance because the cavity effect induced by the atoms is not strong enough, or atomic coherence is not long enough, to affect the spontaneous emission rate. The enhanced photoluminescence is because of \color{black} the reduction in scattering loss imposed by the Bragg resonance. The scattering loss from the ensemble is approximately given by \color{black} $(1+2\beta_{eff})/(1+\beta_{eff})^2$ \color{black}, for an optically thin ensemble. In an optically thick sample, multiple reflections give rise to radiation trapping and scattering loss, which can still be significantly less than the case of random atomic distribution\cite{DEUTSCH:1995aa}.\color{black}

To experimentally study the effects described above, we fabricate microring resonators using a 500nm thick SiN layer on SiO${_2}$. The width and diameter of the ring is 1.5 $\micro$m and 150$\micro$m, respectively. Er ions were implanted into an ion array, separated by an integer number of the Er emission wavelength, with a precision of $\pm$ 10nm at the Sandia National Laboratories. The energy used for the implantation was 200 keV with implantation depth of 50nm. About $10^4$ ions were implanted in a rectangular area of width 20nm along the radius of the ring. The sample was then annealed at 1100$^{\degree}$C for one hour in nitrogen flow of 5.0 standard liter per minute (SLPM). A polymer (PMMA) layer of around 1 $\micro$m thick was spin coated on the sample as an upper-cladding after annealing for better mode confinement and also to fine tune the effective index of the optical mode by varying the thickness of this layer. The sample was placed inside a cryostat with typical temperature of 4K with two optical fibers directing light to/from the sample from/to the room-temperature setup outside the cryostat for measurement. The atoms were resonantly excited by a tunable diode laser. Typical input power of the pump in the experiments was around 5 $\mu W$, below the saturation limit of the erbium ions. Two AOMs were used to create a pulse to pump the atoms for 6ms and turn it off for 10-15ms during the photoluminescence (PL) measurement. The PL light was detected using a single photon detector and counts were averaged over $10^{5}$ runs. 

  \begin{figure*}[!t]
\vspace{0mm} 
	\centerline{\includegraphics[width=2\columnwidth, angle=0]{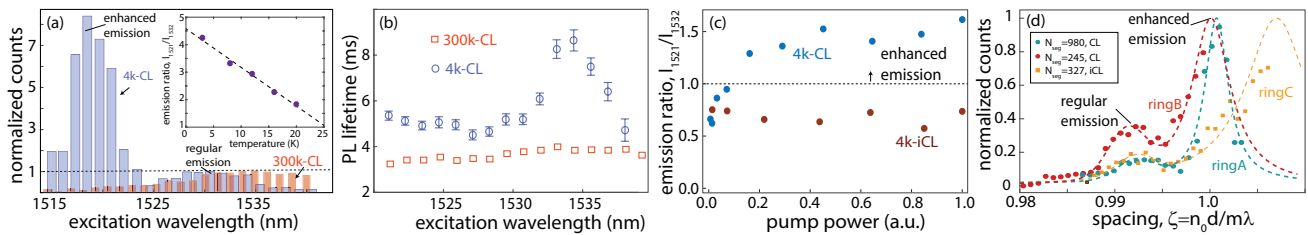}}
	\vspace{0mm}
	\caption{ (a) Resonance photoluminescence spectrum at 4K (blue bars) and 300K (red bars). The enhanced emission around 1520nm is a signature of the \color{black} Bragg resonance \color{black} induced by the atomic lattice.    Inset shows the  ratio between the peak emission photon number at 1521nm (near the \color{black} near the Bragg resonance \color{black}) to that at 1532 nm (near the typical Er emission) as a function of temperature. \color{black} Inset shows the temperature dependency of the emission ratio with a dashed line fitted to guide the eye. \color{black} (b)The PL lifetime is shown for 4K (blue circles) and 300K (red squares). The longer PL decay time around 1535nm is associated with enhanced reabsorption away from the CL condition.  (c) The ratio between the peak emission photon number at 1521nm to that at 1532 nm as a function of pump laser power. Here, for the iCL (ringC), near 1520nm, the emission ratio does not depend on laser power, but for the CL (ringB) the ratio increases with laser power and then saturates. Such nonlinear response across the emission spectrum in the CL is expected when emission  exceeds the re-absorption probability in the CL.	(d) Normalized emission (probability) as a function of spacing parameter, $\zeta$, is plotted for ringA, ringB, and ringC (see text for details). Here ringA and ringB satisfy the CL condition near 1520nm while ringC does not. The dashed lines are Gaussian and Lorentzian fits to the regular and \color{black} Bragg resonance \color{black} emission profiles, respectively.  }
	\label{fig: fig2}
\vspace{-3mm}\end{figure*} 

  %  : change A B C in the figure and text to (a), (b), (c)
  For measuring the emission spectrum, we test three ring resonators, ringA, ringB and ringC with 980, 245 and 327 atomic segment numbers, respectively. The bus waveguide is designed to maximally transmit the TM polarization by tapering the waveguide at the fiber-waveguide interface. Due to the large width of the resonator, TM1 mode is efficiently coupled to the resonator with more than 60\% coupling efficiency having a loaded Q factor of about $7\times10^4$. The calculated effective index for ringA and ringB for this mode is about 1.58. Considering this effective  index, the Bragg resonance condition is satisfied at about 1520nm when the atomic lattice spacing is an integer multiple of \color{black} $1520nm/n_{eff}=962$nm. \color{black} We refer to this as ``commensurate lattice (CL)" at around 1520nm. For ringC, we removed the PMMA layer on the SiN ring resonator. So, the calculated effective index for ringC  for this mode is about   $n_{eff}=1.568$  , which, in theory, shifts the resonance condition to lower wavelengths    ($\sim$ 1508nm).  We refer to this as ``incommensurate lattice (iCL)" condition   where the emission wavelength centered within the Er emission band does not match the lattice spacing.  The lowest wavelength that could be reached with  our laser is 1510nm.  We also note that the implantation of Er ions and annealing conditions for the CL and iCL samples were done under the same conditions and the only difference is the PMMA upper cladding layer and thus the observed effect can not be explained by presence of implantation damage\cite{sobolev1998dislocation, kenyon2005erbium}. The fact that host is amorphous and emission shows pump-power dependency and long-lifetime (corresponding to erbium ions) confirms that the \color{black} enhanced \color{black} emission is not caused by defects in the hosts.\\
 The emitted mean photon number is measured at every free-spectral range of the cavity. The resulted emission spectrum at 4K temperature is shown in Fig.\ref{fig: fig2} (a). The spectrum significantly deviates from the emission of the ions at room temperature. The observed peak emission near 1518nm corresponds to a commensurate lattice with 962nm spacing between implanted segments. Assuming an effective index of 1.578, the result is in agreement with the calculated resonant wavelength value.\color{black} Taking into the account our experimental parameters including the spatial atomic distribution width (20nm), inhomogeneous broadening (3nm), total atom number ($10^7$), yield factor of the implantation (10\%),  branching ratio in Er ions (0.1), and ion-cavity field overlap (0.5), the estimated \color{black}$\beta_{eff}$ \color{black}is about 50. The scattering loss in this limit can be numerically calculated where signifiant enhancement is expected\cite{DEUTSCH:1995aa}. The result in Fig.\ref{fig: fig2} (a) shows an enhancement of >30, in broad agreement with the Bragg theory. \color{black}\\
  As temperature increases the PL ratio decreases suggesting suppression of \color{black} the emission near the Bragg resonance \color{black}. This is because of increase in \color{black} homogenous broadening \color{black} and phonon-assisted excitations~\cite{miyakawa1970phonon} \color{black} (confirmed by observation of increased absolute emission intensity with temperature rise) \color{black} at elevated temperatures. The emission ratio is plotted versus temperature as an inset to Fig\ref{fig: fig2}.(a). The expected \color{black} homogenous linewidth, $\gamma_h$, \color{black} in amorphous hosts like silicon nitride is expected to \color{black} exceed the cavity linewidth, $\kappa$, \color{black}  at higher temperatures\cite{Gong:optexp2010} \color{black} causing $\beta_{eff}$ to decrease\color{black}. Moreover, presence of phonon-assisted excitation \color{black} and also increased inhomogeneous broadening contribute to \color{black}  the reduction of the $\beta_{eff}$ as the temperature increases. Consequently, the spatial interference and \color{black} the Bragg resonance \color{black} effect from the atoms is suppressed    \color{black} at elevated temperatures. It should be noted, that even though we observe enhanced  Bragg effect from the ensemble at low temperatures, the single-atom cooperativity is not large enough to alter the atomic decay rate. \color{black}\\
 The emission decay time measured at different wavelengths (Fig.\ref{fig: fig2} (b)) shows \color{black} lengthening of the lifetime at low temperatures, which we associate to suppression of non-radiative decay. \color{black} The lengthened decay time around 1535 nm is due to the reabsorption \color{black} \cite{SUMIDA:1994aa} (radiation trapping) \color{black} at the peak (regular) emission wavelength where more atoms exist.  The negligible lifetime change around 1520nm ruling out the possibility of dominant superradiance emission from the ions near the atomic Bragg resonance condition.\\ 
 We plot light intensity at around 1521nm ( \color{black} emission near the Bragg resonance \color{black} ) relative to that of 1532nm (near regular emission peak). Figure 2 (c) shows that beyond some pump power, the emission overcomes the re-absorption and \color{black} enhanced \color{black} emission is evidenced. Such enhanced emission is absent for the iCL as the emission signal over the entire spectrum linearly changes with the atom number. At high pump powers, the PL eventually saturates for both transitions.  \\ 
 Figure \ref{fig: fig2}(d) shows the total photons emitted from the ions in ringA, ringB and ringC as a function of the spacing parameter,   $\zeta=n_0d/m\lambda$, where $n_0 =1.58$. As ringA and ringB consists of a CL at around 1520nm, we see the enhanced emission near that wavelength.    Whereas, for ringC, the Bragg resonance is shifted to lower wavelengths.    The center peak ($\zeta\sim 0.992 $) corresponds to the regular emission around 1532 nm, which is inhomogeneously broadened. We use a Gaussian function to model the emission centered around 1532nm with a     full    inhomogeneous width of  about   $3$nm.    The linewidth of the enahnced emission peak is governed by atomic spatial distribution (with a Gaussian width of around 40nm), inhomogeneous broadening of the atomic transitions (around 3nm) and number of atoms. We note that although the number of ions for ringC is more than ringB, the effective atom number, taking into the account the Er atomic transition, is less for ringC giving rise to its broader  emission peak.    By comparing the fitted amplitudes of the Gaussian (describing PL emission around 1532nm) and Lorentzian distribution  (describing the \color{black} enhanced \color{black} emission around 1520nm) functions we observe nonlinear dependence of the relative emission to the atom number. For an iCL, the ratio between the two emission peaks (around 1520nm and 1532nm) is independent of the atom number.   This is evidenced in Fig. \ref{fig: fig2}(c) where power is a proxy for atom number. We observe that the ratio between the \color{black} emission near the Bragg resonance \color{black} and regular PL emission  scales nonlinearly with the number of atomic segments. In the case of data in Fig. \ref{fig: fig2}(d), the resonator with $m=1$ (ringA) has four times more atoms compared to the resonator with $m=4$ (ringB). Using the values extracted from the model plotted in Fig. \ref{fig: fig2}(d) we infer a ratio of about 2.4 between the \color{black} emission near the Bragg resonance \color{black} and the regular emission.   

 %%%%%%%
\begin{figure}[!t]
\vspace{4mm}
	\centerline{\includegraphics[width=0.8 \columnwidth, angle=0]{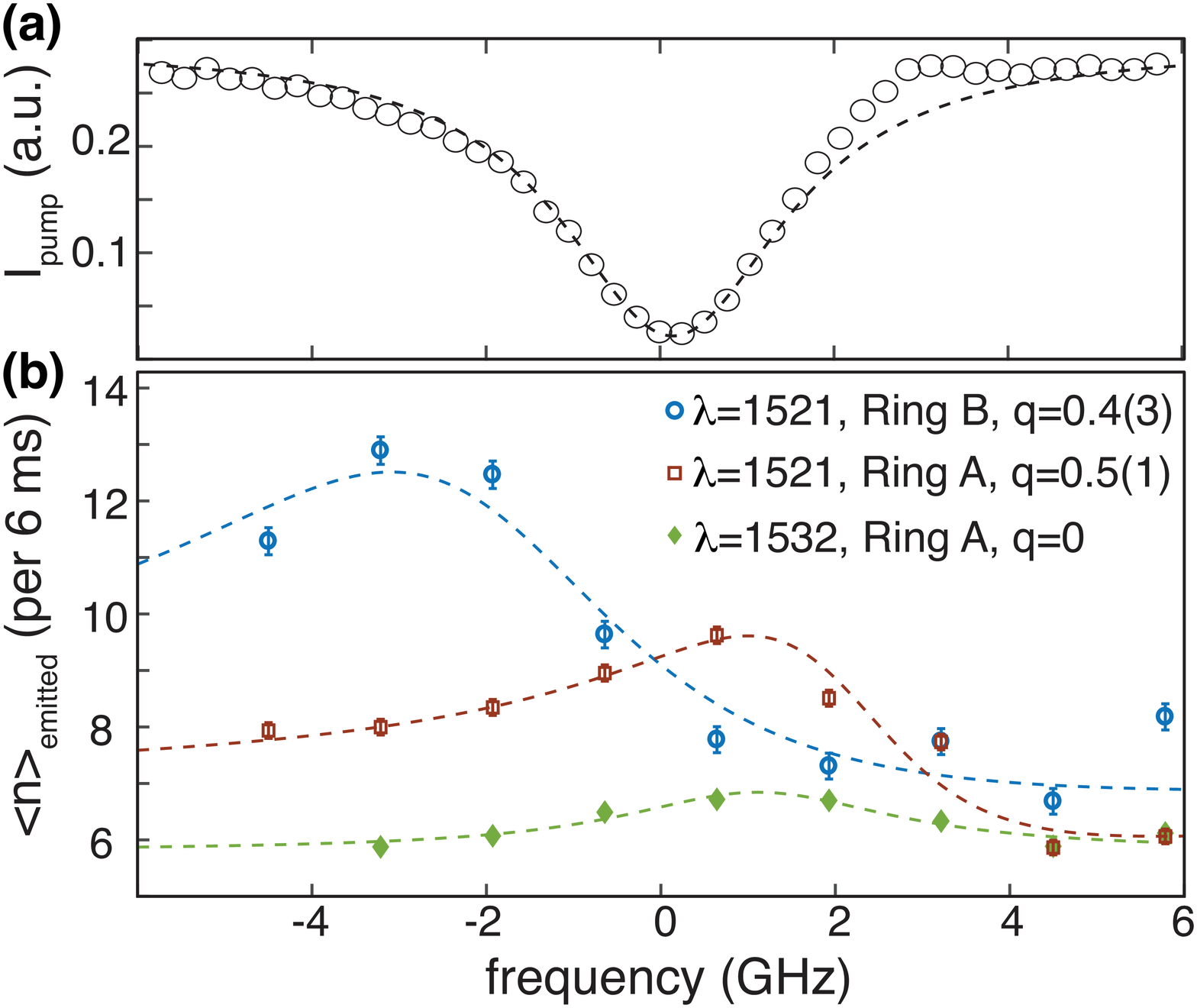}}
	\caption{(a) Transmitted spectrum of the pump light through ringA with fitted Lorentzian width of $2.6\pm0.1$GHz. (b) Emission spectrum near cavity resonance at 1520 and 1532nm for ringA and ringB. The spectrum is fitted with Fano-Lorentzian lines with asymmetry described by Fano parameter. The asymmetry in the lineshape of (a) is due to a change in the rate of the scan which does not appear when a wide scan is used. In the case of (b) the laser frequency is fixed for each point and the lineshape around 1532nm (green data) shows a symmetric Lorentzian, as expected from an iCL.}
	\label{fig: fig3}\vspace{-3mm}
\end{figure}
 	
 Moreover, the presence of atomic Bragg  condition suggests the possibility of mode interference and appearance of Fano-type resonances in the system\cite{limonov2017fano}. In our system, the coupling between the atomic Bragg mode (continuum mode) and the ring resonator (discrete mode) gives rise to asymmetric emission lines, where phase change of the atomic  mode happens very slowly compared to the ring resonator mode. The asymmetry in the lineshape is characterized by the Fano parameter $q=\cot \delta $, where $\delta$ is the phase shift of the atomic  mode with respect to the cavity mode. This is evidenced by the observed lineshapes in Fig. \ref{fig: fig3}(a) and \ref{fig: fig3}(b). A non-zero Fano parameter is fitted to the emission spectrum centered around 1520nm for both ringA and ringB while the emission spectrum around 1532nm (green data) is described by a symmetric Lorentzian function. To obtain this data, we collect PL from the ions after pump excitation at different frequencies around a single cavity resonance near 1520~nm and 1532~nm.      
 
 % conclusion

In conclusion, we design atomic geometries of erbium ions inside a silicon nitride micro-photonic resonator and study the effect of geometry on light-atom coupling from the ensemble. We observe that spatial interference at low temperatures from a lattice of ions creates a Bragg atomic resonance enhancing the \color{black} emission. \color{black} The interference between the optical and atomic \color{black} Bragg \color{black} modes is also observed through Fano-type resonance features.
The investigation offers a unique path to further engineer active silicon photonic structures for  studying cQED interactions between atoms and photons in an ensemble \cite{Chang:2018aa, Chang:2019aa}, on-chip  photon generation\cite{ Barik:2018aa, furuya2020study}, \color{black} and engineering non-Hermitian Hamiltonians  \cite{Feng:2014ab}.  \color{black}

%\subsection{Funding}
\textbf{Funding.} Tellabs Foundation; Center for Integrated Nanotechnologies; U.S. Department of Energy (DE-NA-0003525); Sandia National Laboratories .

\textbf{Acknowledgment.} The views expressed here do not necessarily represent the views of DOE or the U. S. Government. M.H. thanks A. Weiner and C.-L. Hung for the enlightening discussions.

%\section*{Data availability}
%   The data that support the findings of this study are available from the corresponding author upon request.
% \section*{Acknowledgement}
\newpage
\bibliography{allref} 
% Full bibliography added automatically for Optics Letters submissions; the following line will simply be ignored if submitting to other journals.
% Note that this extra page will not count against page length
%\bibliographyfullrefs{allref}

\end{document}